\documentclass{epl}

\title{Local excitations in mean field spin glasses}
\author{Florent Krz{\c{a}}ka{\l}a\inst{1} and Giorgio Parisi\inst{1,2}}%
\institute{
  Dipartimento di Fisica$^{1,2}$, INFM$^{1,2}$, SMC$^{1,2}$, and INFN$^{2}$\\
  Universit\`a di Roma {\em La Sapienza}  P. A. Moro 2, 00185 Roma, Italy
}

\begin{document}

\maketitle

\begin{abstract}
We address the question of geometrical as well as energetic properties of
local excitations in mean field Ising spin glasses. We study analytically the Random Energy Model and numerically a dilute mean field model, first on tree-like graphs, equivalent to a replica symmetric computation, and then directly on finite connectivity random lattices. In the first model, characterized by a discontinuous replica symmetry breaking, we found that the energy of finite volume
excitation is infinite whereas in the dilute mean field model, 
described by a continuous replica symmetry breaking, it slowly decreases with sizes and saturates at a finite value, in contrast with what would be naively expected. The geometrical properties of these excitations are similar to those of lattice animals or branched polymers. We discuss the meaning of these results in terms of replica symmetry breaking and also possible relevance in finite dimensional systems.
\end{abstract}

Spin glasses~\cite{Young98}, disordered magnets displaying many out of equilibrium phenomena such as aging, memory and rejuvenation are, on the theoretical level, now believed to be well described on fully connected and finite connectivity random graphs by the so called replica symmetry breaking (RSB) picture~\cite{MezardParisi87b}, sometimes referred to as the mean field scenario, in which the free energy landscape is divided into many valleys with $O(1)$ energy differences.

For finite dimensional spin glasses, most particularly for the Edwards-Anderson~\cite{EdwardsAnderson75} (EA) model, how to describe the ordered phase still remains controversial. A nice renormalization group picture called the droplet model~\cite{BrayMoore86}, based on real space properties and scaling laws, gives many nice and useful results but it is still not clear if this picture describes well all properties of the EA model. It starts with the very assumption that there is only one state (or ground state at $T=0$, and up to a global spin flip) and that the typical energy of a compact cluster of spins of size $\ell$~\footnote{We define the energy of a cluster as the energy difference between the original configuration and the one where all spins belonging to the cluster are flipped.} goes like  $\ell ^ {\theta}$. The energy distribution of such clusters is assumed to be very large, so the probability of finding a droplet of energy $O(1)$ goes like $\ell ^ {-\theta}$. A positive $\theta$ is then needed to have a finite $T_c$. Compactness and size independence (i.e. that a droplet of size $2 \ell$ is not correlated with the one of size $\ell$) are assumed in this theory. 

There are actually many definitions of $\theta$ in the literature, corresponding to different types of clusters, and it will be of some use here to consider three types of excitations, we will thus need three exponents: (1) $\theta_g$ (g for {\it global}) will describe the energy of low lying system-size excitations~\cite{HoudayerMartin00b} those energies are assumed to scale like $L^{\theta_g}$ where $L^d=N$ is the size of the system, (2) $\theta_l$ (l for {\it local}) will describe the energy of finite volume Minimal Energy Clusters (MEC) of $n$ spins given that a particular spin $i$ belong to it (i.e. for all clusters of $n$ spins that contains the spin $i$, the MEC is the one with minimal energy); their energies scale like $\ell^{\theta_l}$, where $\ell$ is the typical linear size of such a cluster of $n$ spins, and (3) $\theta$, the original droplet exponent, will describe the energy of droplets (they are defined like MEC with the constraint of being compact and size independent, see \cite{BrayMoore86} for a description of the droplet/scaling theory and effective definition of {\it compatness} and {\it size independance}) and their energy goes like $\ell^{\theta}$. This latest droplet exponent is usually computed via domain-wall (DW) studies~\cite{McMillan,Aspel} though it is not straightforward that droplets and domain walls share the same energetic properties, i.e. that $\theta = \theta_{DW}$. A natural realization of this droplet picture arises in spin glasses on hierarchical lattices, or in the Migdal-Kadanoff renormalization group(MKRG), where one can compute analytically energetic properties and show that $\theta_{DW}=\theta=\theta_l=\theta_g$~\cite{BouchaudKrzakala02} i.e. that domain-walls, droplets, MEC, and low energy system-size excitations have the same energetic (a typical energy growing like $\ell^{\theta}$) and geometrical (they are compact with a fractal surface dimension $d_s=d-1$) properties.

 In mean field systems however, there are several macroscopically different states so that it must be possible to find (at least) system-size (involving a finite fraction of spin of the system) excitations of $O(1)$ energy with a {\it finite} probability. In our notation, this many valley picture thus implies $\theta_g=0$ (it has also been suggested that these system-size objects exist in the EA model; they are called sponges in this context due to their topological properties and 
according to mean field predictions~\cite{HoudayerMartin00b} they have $O(1)$ energy and a fractal surface dimension $d_s$ equal to the space dimension $d$). For finite volume excitations, i.e. for the very equivalent of droplets and MEC of $n$ spins (or of typical size $\ell$) in mean field systems, the situation is unfortunately still rather unclear. It was first suggested that in mean field models these MEC should be like droplets, i.e. that they have an energy growing with their size~\cite{HoudayerMartin00b} and the probability to find one with energy $O(1)$ should decrease like $\ell ^ {-{\theta}_l}$ (with $\theta_l=\theta>0$) so that only system-size excitations have low energies ($\theta_g=0$) and non trivial properties. The {\it a-priori} reason for such a belief comes the Bethe-Peierls argument stating that if, locally, $\theta \leq 0$ then there is {\it no} spin glass phase. We will show that an ideal realization of such a scenario arises in Derrida's Random Energy Model (REM)~\cite{Derrida80} in which $\theta_g=0$ and $\theta_l=\theta=\infty$. The difference between mean field and scaling pictures would then lie in the value of $\theta_g$.

It has been however noticed~\cite{LamarcqBouchaud02} quite recently that the energy of MEC could be different from droplet's and may decrease with size without jeopardizing the spin glass phase at finite $T$ provided that they are not size independent, i.e. that excitations of size $2\ell$ are not independent of excitations of size $\ell$ and are correlated enough. Hence the value of $\theta_{l}$ (which does not assume independence) might be different from $\theta$ and can {\it a-priori} even be negative; this remains an open question.  In \cite{LamarcqBouchaud02}, the authors found indeed a negative, thought very close to zero, value of $\theta_{l}$ in the 3d EA model; thus it would be nice to have similar results in systems which are known to have a finite temperature phase transition.  Moreover the relevance of mean field results in finite dimensional systems is still very controversial and in order to understand what mean field has to say on this matter, we should understand what properties of local excitations in mean field models really are. It is also useful, for the very understanding of the mean field spin glass phase and of the continuous RSB phenomenon, which has been proved to be useful in many systems over the last few years, to investigate these properties and to construct a sort of droplet picture for these systems. Finally, due to a proliferation of definitions of $\theta$s and numerical studies of their values, it is also useful to fix the notation and to understand links between them. It is the aim of this letter to fill these gaps.

We study in this letter two mean field Ising models: (1) the REM where we find $\theta_g=0$ and $\theta=\theta_l=\infty$, and (2) the finite connectivity Ising spin glass with Gaussian couplings in two different cases: (a) on a Cayley tree with fixed random boundary conditions, a system equivalent to a replica symmetric (RS) approximation where finding ground states and MEC is easy, and (b) on random graphs (or Bethe lattice, see~\cite{MezardParisi01} for a precise definition), the well defined but harder to study, mean field model. In the REM we use simple analytical arguments whereas in finite connectivity models we find ground states numerically, then pick a spin at random and look for the cluster of $n$ spins containing the previous one which minimizes the energy (the MEC); that gives a $E(n)$ that we average over many instances. Our main result is that the MEC energy {\it decreases} with their size and saturates to a finite value, giving $\theta_l=\theta_g=0$ and thus violate the Bethe-Peierls argument (note that we defined in this work mean field $\theta$s exponent by $E=n^{\theta}$ where $n$ is the number of spins rather than with the linear size $\ell$ of the droplet. This make no changes in this study since we find either $0$ or $\infty$). We also find that the geometrical properties of these excitations are those of lattice animals, which gives some hints to the value of the upper critical dimension $d_u$. In conclusion, we comment on these results and their potential relevance in the EA model in the light of~\cite{LamarcqBouchaud02}. 

\section{The Random Energy Model} At zero temperature the REM, as defined as the $p \rightarrow \infty$ limit of p-spin models~\cite{Derrida80}, is a fully connected spin model where all configurations have random energies taken from a Gaussian distribution of mean $0$ and variance $N$. It has a ground state of energy $E_0 \propto - N\sqrt{\ln{2}}$ with variance of $O(1)$. They are also $O(1)$ energy excitations of size $O(N)$ leading to RSB; thus according to our definition we have $\theta_g=0$. Let us consider now finite volume excitations. The number of excited clusters of $n$ spins that contains a particular spin $i$ is $O({N}^{n})$; a straightforward extremal computation~\footnote{The typical minimum of $x$ random values taken from a Gaussian distribution of variance $N$ is $-\sqrt{N\ln{x}}$. The ground state is thus $E_0 \approx -N \sqrt{\ln{2}}$ and the minimum of $N^n$ values is $E(n) \approx -\sqrt{N n \ln{N}}$.} gives that the typical energy of such clusters is $O(-\sqrt{nN\ln{N}})$ and a comparison with ground state energy $E_0$ then gives that energy difference is $O(N)$. We thus obtain $\theta_l=\infty$ for finite volume excitations (i.e. $n << N/\ln{N}$) and only system-size (with $n >> N/\ln{N}$ spins) excitations have $O(1)$ energies and $\theta_g=0$.

The REM is thus an exact realization of the scenario described in~\cite{HoudayerMartin00b}. This has a simple interpretation in terms of one-step RSB in which the energy landscape is dominated by macroscopically different valleys but where, inside a single valley, the energy increases while going away from the bottom. In the REM the valley is composed by only {\it one} configuration, this is why $\theta_l=\infty$, and a natural expectation is that, in more realistic models where there is a similar clustering in the energy landscape, $\theta_l$ should be positive but finite. However the REM is described by a so-called one step RSB ansatz and the situation might be different in models with full-step RSB that we shall now consider.

\section{The spin glass on a Cayley tree} Before considering finite connectivity mean field spin glasses, let us first consider the problem on a tree with fixed boundary conditions. This system is known not to be a very good mean field spin glass and a computation on such lattices is actually equivalent to a RS approximation of the result on random graphs. However, since numerical computations are easier on this type of graph, we will obtain a quasi-thermodynamic (more than $2^{20}$ spins) limit result, free from finite size $N$ effects, that will turn out to be a useful RS approximation of the real value of MEC energies.

\begin{figure}
\includegraphics[width=0.6\textwidth]{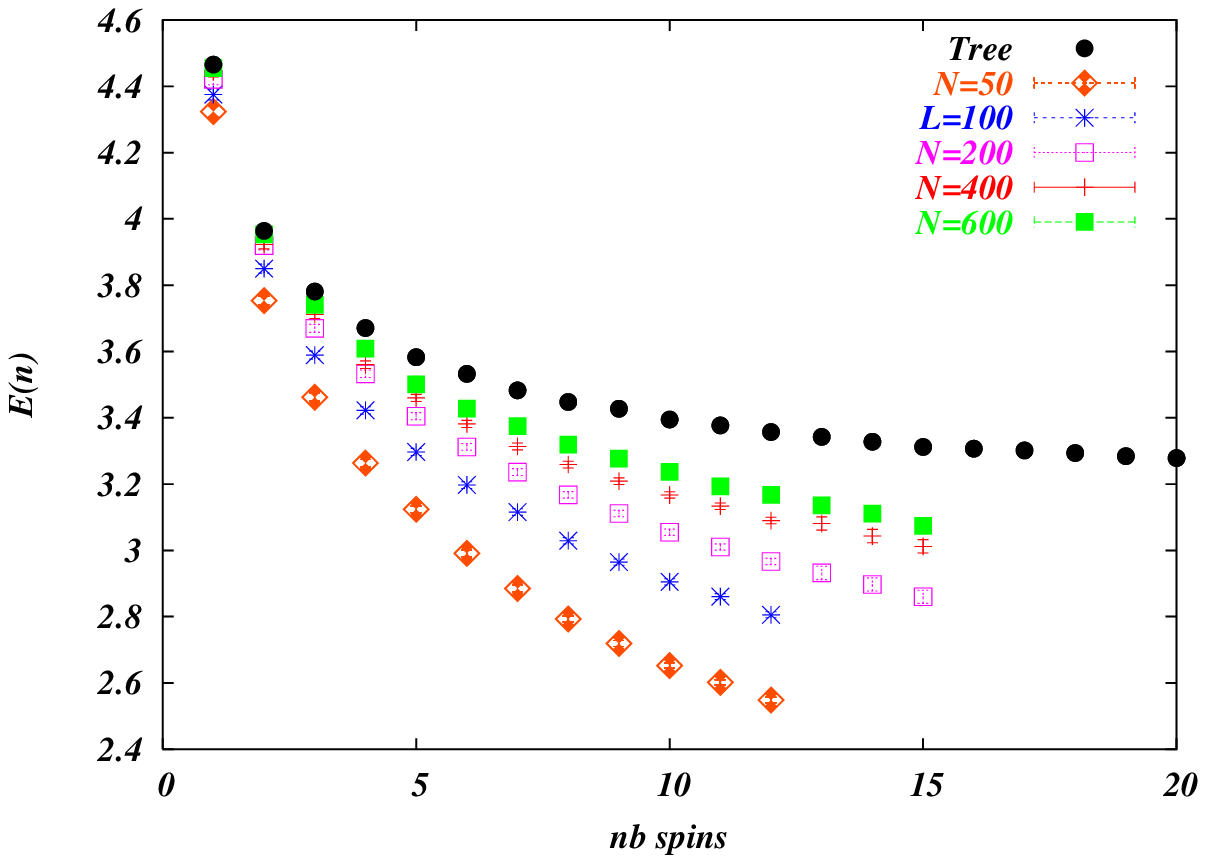}
\includegraphics[width=0.42\textwidth]{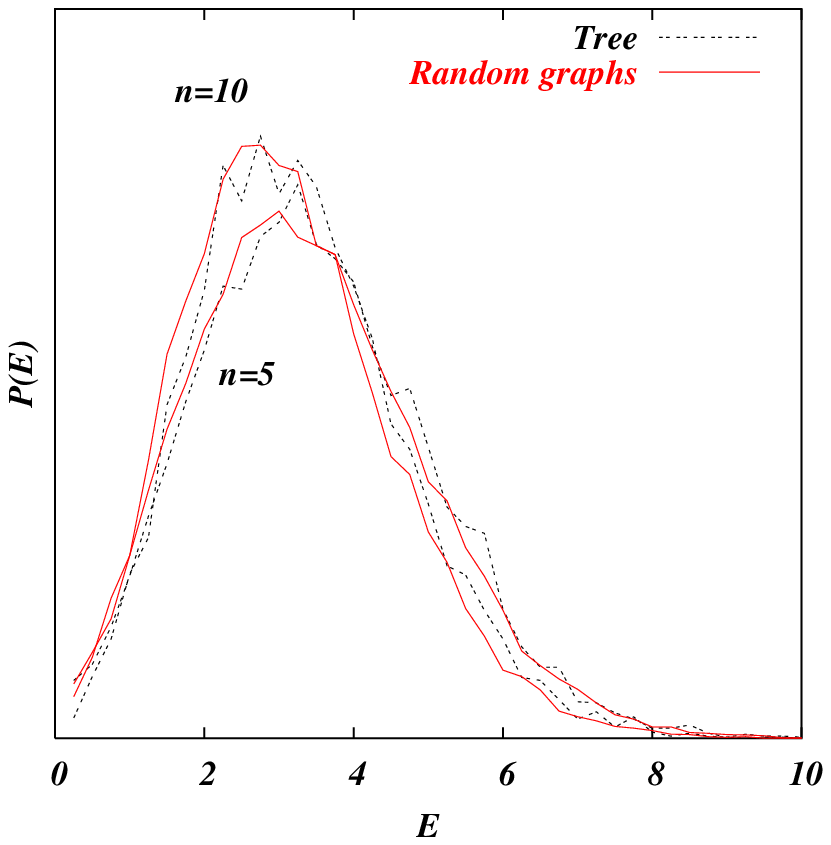}
\caption{Left: mean energy of a MEC of $n$ spins in spin glasses on a tree and on random graphs of different sizes $N$ for connectivity $3$. Tree curve, corresponding to the RS approximation, is believed to be an upper bond. These curves clearly saturates at a finite value implying $\theta_l=0$. Right: distribution of such energies for a MEC of $5$ and $10$ spins on a tree (dashed curves) and on $N=600$ random graphs (full line).}
\label{Energy}
\end{figure}

A Cayley tree of connectivity $k+1$ is built starting from a central site, and constructing a first shell of $k+1$ neighbors. Each of these first shell spins are then connected to $k$ new neighbors in a second shell etc.. until one reaches the $L'th$ shell which is the boundary. Thus a finite fraction of spins are on the boundary, making it a very inhomogeneous system for which properties are dependent on boundary conditions; indeed even in an infinite system we can not forget them since they fix the degree of frustration. This system is known to exhibit a finite $T$ spin glass transition. Finding ground states and excitations on such a tree is a polynomial problem. We are thus able to compute ground states up to very big sizes and average over many instances and thus our results are very accurate. We will focus on $k=2$ (connectivity $c=3$). The energy $E(n)$ of the MEC of size $n$ is found clearly to {\it decrease} and to saturate at large values. A fit of our data gives, with good precision, that the mean energy $E(n)$ is close to $E(n)=A + B n ^{-2/3}$, $A$ and $B$ being positive constants. The exponent close to $-0.66$ is interesting in itself as it has already been noticed to play an important role in ground state energy finite size corrections~\cite{BouchaudKrzakala02,ParisiRitort93a}. When a large magnetic field $B$ is applied, the energy increases with size; for smaller field ($B \leq 0.6$, $B_c$ being $0.48$ in this model~\cite{These}) we could not see this effect and we obtain data similar to those at $B=0$. On fig.~\ref{Field} a finite size scaling of energy difference $\Delta E(n)=E^B(n)-E^{B=0}(n)$ is plotted vs $(B-B_c)\sqrt{n}$, using $B_c=0.48$. Data collapse is excellent~\footnote{$\Delta E(n)$ is a bit negative when $B<B_c$; this seems to be a pathology of the Cayley tree that vanishes at $B>B_c$.}.

Asymptotic energy of MEC seems finite in the RSB phase, thus ${\theta}_l=0$. Since we know that there is a finite temperature phase transition we conclude that mean field spin glasses violate the Bethe-Peierls argument. Moreover, this shows very nicely that a system of size $N+1$ differs from the one of size $N$ because it flips these large system-size clusters so its ground state changes completely from size to size. Fixing boundary conditions also prevents these system-size excitations from being thermally excited, which is why this problem is finally RS with a trivial $P(q)$ and why, in order to have a well defined system, it is better to define a Bethe lattice with random graphs~\cite{MezardParisi01}.

\section{Spin glass on a random graph} 
\begin{figure}
\includegraphics[width=0.6\textwidth]{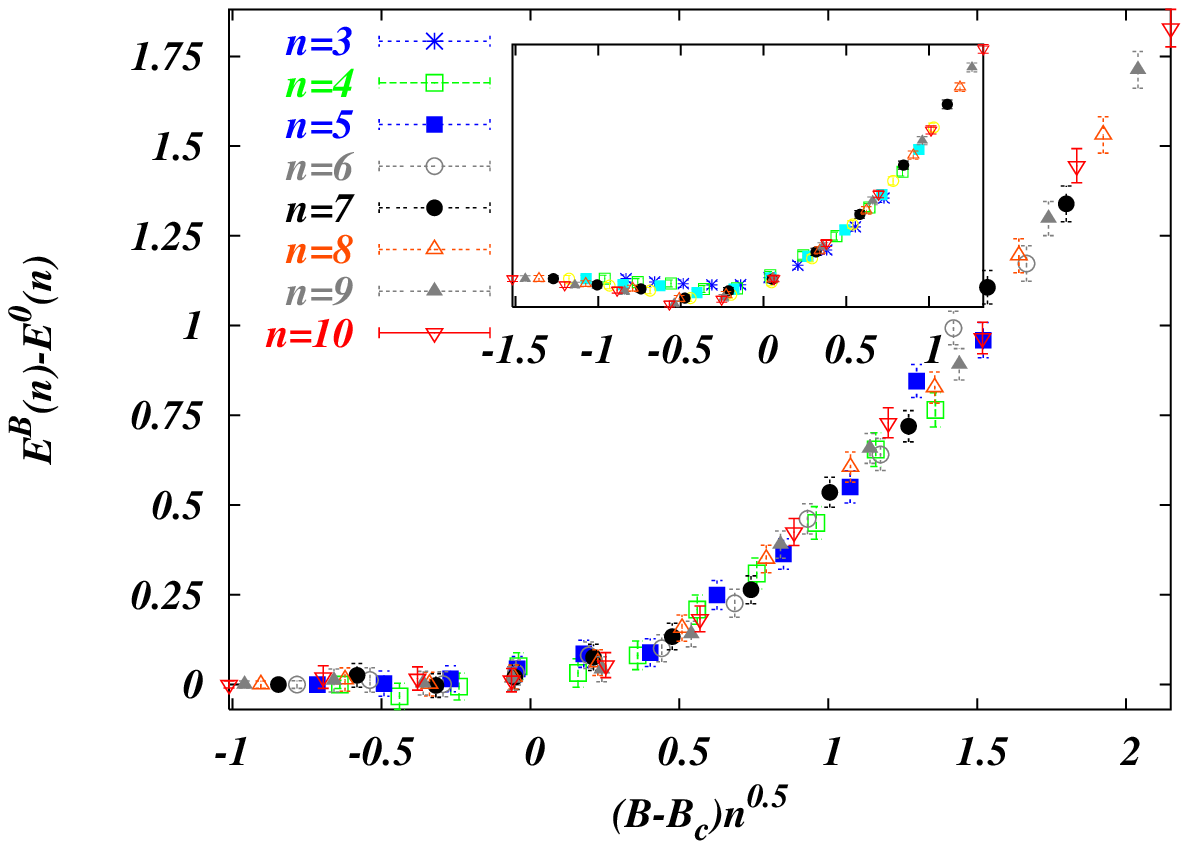}
\includegraphics[width=0.42\textwidth]{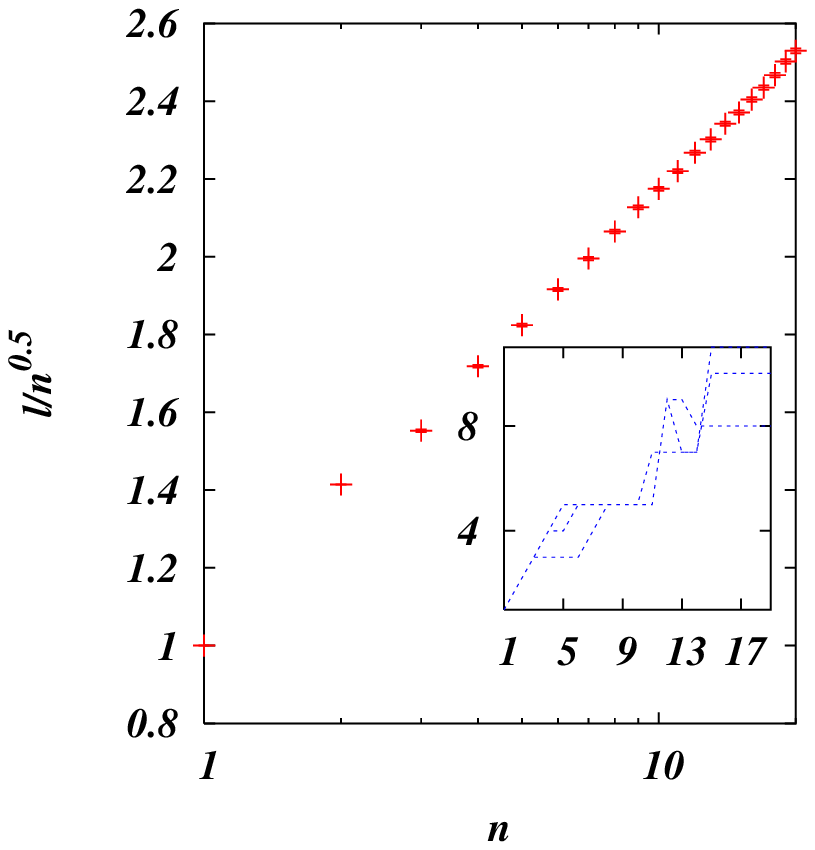}
\caption{Left: rescaled energy difference $\Delta E(n)=E^B(n)-E^{B=0}(n)$ versus $B$ on $N=400$ random graphs with $B_c=0.32$; In inset the same plot on Cayley trees with $B_c=0.48$. Right: end to end distance scale like $\sqrt{n} \ln{n}$ in the spin glass phase (this plot) as well as in the paramagnetic or the disordered ferromagnetic phase, suggesting that these clusters are in a lattice animals phase. In inset, $l(n)$ for $3$ different instances, showing strong correlations among sizes.}
\label{Field}
\end{figure}

Spin glasses on finite connectivity random graphs are supposed to be described by a continuous RSB ansatz; solving their ground states is exponentially hard and now we have to use a heuristic algorithm~\cite{HoudayerMartin01}. We then search MEC of size $n$ by choosing a spin at random and making an exhaustive search of all clusters and finally choosing the one with minimal energy. We can {\it a priori} expect $E(n)$ to have a behavior similar to the RS one because RS approximation usually gives an upper bound to the full RSB behavior, in which case $E(n)$ should not increase with $n$ in this model. MEC energies are plotted on fig.~\ref{Energy} for connectivity $3$ and seems to confirm that RS results are upper bounds. Again, energy decreases and seems to saturate when $n \rightarrow \infty$. We also see a strong $N$ dependence of $E(n)$ that gets closer to the RS result when $N$ increases. When we fit these curves, we obtain a finite asymptotic value that increases with $N$ and an exponent going from $-0.2$ than $-0.4$, compatible with a value $-0.66$ at $N=\infty$. We have checked that all these results remain qualitatively the same when we change some parameters like the distribution of disorder (Gaussian couplings or discrete $\pm J$) or the connectivity (we find similar results for $c=4$,$5$ or $6$, though numbers of spins ($\approx 10$ spins) considered were smaller). Again, if a magnetic field is applied, we are able to reproduce the finite size scaling observed on the tree, though with a smaller value of $B_c$, which is probably due to finite $N$ effects (fig.~\ref{Field}), the $B_c$ giving the best scaling increasing with $N$. Thus, again, $\theta_l=0$ seems true in the whole spin glass phase.

 These results strongly suggest a deep link between the continuous RSB and the presence of low energy excitations of all sizes: it seems to be {\it because} of the presence of such excitations that the replica symmetry {\it has} to be broken. To conclude, mean field spin glass goes against the Bethe-Peierls argument and energy exponents of local as well as global excitations are $\theta_l=\theta_g=0$.

\section{Geometrical properties of excitations} For entropic reasons, we may expect these excitations to be in a lattice animal, or branched polymer, phase, as suggested by~\cite{LamarcqBouchaud02}. On a tree the gyration radius of such clusters is growing like $\sqrt{n}$~\cite{Degennes68,Lub}; it is however easier, for numerical reasons, to compute the average distance $l(n)$ from the root of the cluster of $n$ spins to the deepest point in the tree. In that case, since a lattice animal cluster of size $n$ is more less composed by $\sqrt{n}$ independent branches with typical size $\sqrt{n}$ and a Bose-Einstein like distribution~\cite{WyartBouchaud02}, a simple straightforward computation of extremes gives that $l(n) \propto \sqrt{n}\ln{n}$ for lattice animals~\footnote{The length distribution behaves for largest length like $e^{-l/\sqrt{n}}$, the extremal value $l_{max}$ of $\sqrt{n}$ realizations is then given by $\int^{\infty}_{l_{max}} e^{-l/\sqrt{n}} dl = 1/\sqrt{n}$ thus $l_{max} \propto \sqrt{n} \ln{n}$.}. This scaling works perfectly well (see fig.\ref{Field}). Note also that lattice animals have a upper critical dimension $d_u=8$ which is also the dimension at which mean field spin glass theory break (amazingly the related branched polymer problem has $d_u=6$ in $\theta$-solvent, and $d_u=8$ in the generic case; these two dimensions are known to be also the two where something happens in spin glass field theories~\cite{FisherSompolinsky85}), giving more credit to the link between these excitations and the presence of continuous RSB. To illustrate the size dependence, $l(n)$ is also plotted for $3$ particular instances, showing the correlation between $l(n)$ at different $n$.

\section{Discussion} We have studied properties of local excitations involving a finite number of spins in mean field spin glasses. In the REM, a 1RSB model, we found that the energies of these excitations are infinite, so $\theta_l=\infty$. However we find something different on a Bethe lattice, a full RSB model, where the energy decreases and saturates to a finite value. Though it might re-increase at very large values of $n$, this is unnatural in such graphs where the typical size is unity and we do not see any sign of that; the natural interpretation is thus that $\theta_l=0$ and that there are clusters of all sizes with finite $O(1)$ energy.  Finally, the geometrical properties of these excitations are those of lattice animals (note that somehow similar statements on branched clusters in spin glasses was proposed in~\cite{Garel}) , which may justify why dimensions $6$ and $8$ play such an important role in spin glasses mean field theories. 

	Any minimal scenario of real space excitations for spin glasses with full RSB should then include this branch of finite size clusters, as was first proposed in~\cite{LamarcqBouchaud02}, different from the droplet one, in which MEC behaves with ${\theta}_l = 0$. If there are also system-size (thus involving a finite fraction of the system) $O(1)$ energy excitations, they lead to non-trivial $P(q)$. However this is impossible for a spin glass on a tree because fixing boundary conditions prevents large spin clusters to flip, this is why a spin glass on a tree gives an RS approximation of the real mean field system (i.e on a tree $\theta_l=0$ but $\theta_g>0$). Confirming these results in other connectivities, for larger sizes and other RSB models, as well as investigating differences seen between one-step and continuous RSB, would be very useful. It would be also nice to understand more precisely the link between the properties of these excitations and the instability of the RS and successive RSB approximations within the replica method where full RSB models (in contrast to 1RSB model) are characterized by the presence of Goldstone modes (see for instance~\cite{Cyrano}) suggesting, in full agreement with our work, that there are excitations of all sizes with $O(1)$ energy. 

	There is finally the question of relevance of these excitations in finite $d$. Assuming, with~\cite{BouchaudKrzakala02}, the existence of MEC with similar properties, we can expect in high dimension that they induce a RSB phenomenon although the way MEC turn into sponges is actually not straightforward. Some semi-analytic results suggest how it may happens (there are for instance many ground states in some modified spin glass models in $d>8$~\cite{NewmanStein94}), but it is not obvious to see how energetic and geometrical properties of MEC should be modified when they turn into space spanning objects and start to wraps around the periodic boundary conditions. In lower dimension the situation is even more matter of debates, and some effects below $d=8$ might also induce deviations from mean field behavior. However, it may be worth noticing that (a) the usual {\it droplet} branch may be quite well described by the MKRG, as suggested by many results and equalities between {\it a-priori} unrelated exponents~\cite{BouchaudKrzakala02} and the fact that lower critical dimensions are well predicted by these approaches, (b) the {\it mean field} branch of excitations may also exist in these models, as argued in~\cite{LamarcqBouchaud02} who finds numerically some lattice-animals like MEC with $\theta_l \leq 0$ in $3d$ (they seem however to have larger energies and a positive exponent when a magnetic field is applied, which may indicate the absence of spin glass phase in a magnetic field, in contrast to the present study) with a $P_n(E)$ very similar to ours (see however~\cite{Moore} for debates on these results) and (c) there is also much numerical evidence for $\theta_g \approx 0$ in $3d$~\cite{KrzakalaMartin00}. 

\acknowledgments
We thank J-P.~Bouchaud and O.C.~Martin for very useful discussions. This work was supported by the European Community's Human Potential Program under contract HPRN-CT-2002-00319 (STIPCO).

\bibliographystyle{prsty}
\bibliography{../../Bib/references}

\addcontentsline{toc}{chapter}{\protect\bibname}
\begin{thebibliography}{10}

\bibitem{Young98}
{\em Spin Glasses and Random Fields}, edited by A.~P. Young (World Scientific,
  Singapore, 1998).

\bibitem{MezardParisi87b}
M. M{\'e}zard, G. Parisi, and M.~A. Virasoro, {\em Spin-Glass Theory and
  Beyond} (World Scientific, Singapore, 1987).

\bibitem{EdwardsAnderson75}
S.~F. Edwards and P.~W. Anderson, J. Phys. F {\bf 5},  965  (1975).

\bibitem{BrayMoore86}
A.~J. Bray and M.~A. Moore,  in {\em Heidelberg Colloquium on Glassy Dynamics},
  Vol.~275 of {\em Lecture Notes in Physics}, edited by J.~L. van Hemmen and I.
  Morgenstern ({S}pringer, {B}erlin, 1986), pp.\ 121--153. D.~S. Fisher and D.~A. Huse, Phys. Rev. Lett. {\bf 56},  1601  (1986).

\bibitem{HoudayerMartin00b}
J. Houdayer and O.~C. Martin, Europhys. Lett. {\bf 49},  794  (2000). J. Houdayer, F. Krzakala, and O.~C. Martin, Eur. Phys. J. B. {\bf 18},  467 (2000).

\bibitem{LamarcqBouchaud02}
J. Lamarcq, J.-P. Bouchaud, O.~C. Martin, and M. M\'ezard, Europhys. Lett. {\bf
  58},  321  (2002). J. Lamarcq, J.-P. Bouchaud, and O.~C. Martin, Phys. Rev. B 68, 012404 (2003).

\bibitem{McMillan}
W. L. McMillan, J. Phys. C 17, 3179 (1984). 

\bibitem{Aspel}
T. Aspelmeier et al., Phys. Rev. Lett. 90, 127202 (2003).

\bibitem{BouchaudKrzakala02}
J.-P. Bouchaud, F. Krzakala and O.C. Martin, Phys. Rev. B {\bf 68}, 224404 (2003).

\bibitem{Derrida80}
B. Derrida, Phys. Rev. Lett {\bf 45},  79  (1980). D. Gross et al, Nucl. Phys. B {\bf 240},  431  (1984).

\bibitem{MezardParisi01}
M. M{\'e}zard and G. Parisi, Eur. Phys. J. B {\bf 20},  217  (2001) and
J. Stat. Phys 111 (2003).

\bibitem{ParisiRitort93a}
G. Parisi, F. Ritort, and F. Slanina, J. Phys. A {\bf 26},  247  (1993). S. Boettcher, cond-mat/0208196.

\bibitem{These}
A. Pagnani, G. Parisi and M. Ratieville, Phys. Rev. E 68, 046706 (2003).

\bibitem{HoudayerMartin01}
J. Houdayer and O.~C. Martin, Phys. Rev. E {\bf 64}, 056704  and Phys. Rev. Lett. {\bf 83}, 1030 (2001).

\bibitem{Degennes68}
P. de~Gennes, Biopolymers {\bf 6},  715  (1968). 

\bibitem{Lub}
T. Lubensky and J. Isaacson, Phys. Rev. A {\bf 20},  2130  (1979).

\bibitem{WyartBouchaud02}
M. Wyart and J.~P. Bouchaud, cond-mat/0210479.

\bibitem{FisherSompolinsky85}
D.~S. Fisher and H. Sompolinsky, Phys. Rev. Lett. {\bf 54},  1063  (1985). C. de~Dominicis, I. Kondor, and T. Temesvari,  in [1].  C. de~Dominicis and T. Temesvari, Phys. Rev. Lett. {\bf 89}, 097204 (2002).

\bibitem{Garel}
T. Garel, Physics Letters 99A, 402 (1983).

\bibitem{Cyrano}
C. De Dominicis et al J. Phys IV 13 (1998), Eur.Phys.J.B 11, 629-634 (1999).

\bibitem{NewmanStein94}
C.~M. Newman and D.~L. Stein, Phys. Rev. Lett. {\bf 72},  2286  (1994).

\bibitem{Moore}
A.~K. Hartmann and A. P. Young, Phys. Rev. B 66, 094419 (2002).
A.~K. Hartmann and M.~A. Moore, Phys. Rev. Lett. 90, 127201 (2003). 
L.~Berthier and A.~P. Young, cond-mat/0304576.

\bibitem{KrzakalaMartin00}
F. Krzakala and O.~C. Martin, Phys. Rev. Lett. {\bf 85},  3013  (2000).
M. Palassini and A.~P. Young, Phys. Rev. Lett. {\bf 85},  3017  (2000).
E. Marinari and G. Parisi, Phys. Rev. Lett. {\bf 86},3887 (2000).

\end{thebibliography}

\end{document}